• Article •

# Contrasting physical properties of the trilayer nickelates Nd$_4$Ni$_3$O$_{10}$ and Nd$_4$Ni$_3$O$_8$

Qing Li, Chengping He, Xiyu Zhu†, Jin Si, Xinwei Fan, and Hai-Hu Wen*

*Center for Superconducting Physics and Materials, National Laboratory of Solid State Microstructures and Department of Physics, National Collaborative Innovation Center of Advanced Microstructures, Nanjing University, Nanjing 210093, China*

We report the crystal structures and physical properties of trilayer nickelates Nd$_4$Ni$_3$O$_{10}$ and Nd$_4$Ni$_3$O$_8$. Measurements of magnetization and electrical resistivity display contrasting behaviors in the two compounds. Nd$_4$Ni$_3$O$_{10}$ shows a paramagnetic metallic behavior with a metal to metal phase transition($T^*$) at about 162 K, as revealed by both magnetic susceptibility and resistivity. Further magnetoresistance and Hall coefficient results show a negative magnetoresistance at low temperatures and the carrier type of Nd$_4$Ni$_3$O$_{10}$ is dominated by hole-type charge carriers. The significant enhancement of Hall coefficient and resistivity below $T^*$ suggest that effective charge carrier density decreases when cooling through the transition temperature. In contrast, Nd$_4$Ni$_3$O$_8$ shows an insulating behavior despite small value of resistivity at room temperature. The compound shows paramagnetic behavior, with the similar magnetic moments as in Nd$_4$Ni$_3$O$_{10}$ derived from the Curie-Weiss fitting. This may suggest that the magnetic moments in both systems are contributed by the Nd ions. By applying pressures up to about 49 GPa, the insulating behavior is still present and becomes even stronger. Our results suggest that the different Ni configurations (Ni$^{1+/2+}$ or Ni$^{2+/3+}$) and competition between localized and itinerant electrons may account for the contrasting behaviors in trilayer nickelates Nd$_4$Ni$_3$O$_{10}$ and Nd$_4$Ni$_3$O$_8$.

## 1 Introduction

Transition metal oxides with quasi-two-dimensional layered structures often attract special attention in searching for superconductors with high critical temperature (high-$T_C$ SCs). The inspiration comes from the discovery of high temperature superconductivity in the copper oxide perovskite La$_{2-x}$Ba$_x$CuO$_4$ [1] by Bednorz and Müller. Later efforts have led to some success, for example, superconductivity with unconventional properties has been observed in 3$d$/4$d$ transition metal oxides Na$_x$CoO$_2$·yH$_2$O(x∼0.35, y∼1.3) [2], Sr$_2$RuO$_4$ [3] and so on. To find new high-$T_C$ SCs, one strategy is to design materials in which electronic and structural features are analogous to cuprates [4, 5]. Nickel-based oxides have been studied for a long time since nickel and copper are adjacent elements in the periodic table. Theoretical and experimental efforts have been done to search superconductivity in those materials, such as mixed-valent Ni$^{1+}$/Ni$^{2+}$ nickelates and LaNiO$_3$/LaMO$_3$ heterostructures [6, 7]. Recently, the observation of superconductivity in Sr-doped NdNiO$_2$ films have been reported [8] and many theoretical works have followed up on this issue [9–15]. Although the observation of superconductivity in bulk samples is still lacking up to now [16], the related investigation on layered nickelates, especially on the materials containing $Ni^{1+}(3d^9)$ charge state, becomes an important research topic in condensed matter physics [12–15].

---

*Corresponding authors (Xiyu Zhu, email: zhuxiyu@nju.edu.cn; Hai-Hu Wen, email: hhwen@nju.edu.cn).

Rare-earth nickelates with the chemical formula $R_{n+1}Ni_nO_{3n+1}$ ($n$ = 1, 2, 3, and ∞) belong to the Ruddlesden-Popper (*RP*) series compounds [17, 18]. The structure can be described as the stacking of perovskite blocks $(RNiO_3)_n$ and rock salt (*R-O*) layers. In general, the electronic properties of *RP*-type nickelates are highly dependent on the $n$ values and oxygen nonstoichiometry [19–22]. Among them, the $n$ = 3 compounds, namely $R_4Ni_3O_{10}$ ($R$ = La, Pr, and Nd; denoted as: *R*4310) [23–25], have attracted much attention due to the metallic ground state and density-wave-like transition at intermediate temperatures. *R*4310 have been reported to undergo a metal to metal transition at temperature varying from 140 K to 165 K with an pronounced peak or kink feature from specific heat, resistivity, and magnetization measurements [21, 22, 26–28]. The phase transitions ($T^*$) [29] in *R*4310 have been attributed to the charge-density-wave instability [30, 31], but it remains an open question, and more experimental evidence are needed to distinguish the density-wave-like transition in *R*4310. Moreover, another layered nickelate phase $R_4Ni_3O_8$ ($R$ = La, Pr, and Nd; denoted as *R*438) can be obtained by oxygen reduction from perovskite *R*4310 phase [6, 32–36]. Each unit cell of *R*438 (Tetragonal; space group: *I4/mmm*) consists of three square $NiO_2$ planes separated by $R^{3+}$ cations which results in a structural arrangement similar to that of some cuprates, like $La_2Ca_2Cu_3O_8$ [37]. Unlike *R*4310, the physical properties of *R*438 vary greatly with different rare-earth elements. For example, $La_4Ni_3O_8$ exhibits insulating behavior and undergoes a semiconductor-insulator transition at about 105 K [35]. And NMR study reveals the presence of antiferromagnetic order below 105 K [38]. Some theoretical works have been proposed to explain the electronic and magnetic behaviors of $La_4Ni_3O_8$, but the origin is still unsettled [6, 35, 39, 40]. However, another *R*438 series compound $Pr_4Ni_3O_8$ was reported to have a metallic ground state without the feature of phase transition from crystal structure and specific heat measurements [32]. Nonetheless, the detailed studies on other *R*438 type compounds, such as $Nd_4Ni_3O_8$ is still lacking. More importantly, the average valence of Ni in *R*438 is +1.33, which is very close to the valence state of Ni in the reported superconducting nickelate $Nd_{1-x}Sr_xNiO_2$ thin films [8] and cuprates. Thus, it is interesting to know whether the superconductivity can also be realized in *R*438, especially $Nd_4Ni_3O_8$.

In this paper, we focus on the structures and physical properties of the trilayer nickelates $Nd_4Ni_3O_{10}$(Nd4310) and $Nd_4Ni_3O_8$(Nd438). By using a high temperature solid state reaction and subsequent low temperature topochemical reduction method, we successfully synthesize the title compounds. The magnetic susceptibilities of both samples show a Curie-Weiss behavior. In Nd4310, hole-type charge carriers are confirmed from Hall effect measurement and a significant reduction of carrier concentration below the transition temperature ($T^*$) around 162 K is observed. However, Nd438 exhibits an insulating behavior without any clues of phase transition, which is different from the reported $La_4Ni_3O_8$ and $Pr_4Ni_3O_8$ [32, 35]. The high-pressure study on electrical resistance indicates there is no superconductivity emerging up to 49 GPa above 2 K, but an evident enhancement of insulating behavior under high pressures is observed.

## 2 Materials and methods

Polycrystalline samples of $Nd_4Ni_3O_{10}$ were synthesized by a two-step solid state reaction at high temperatures with flowing oxygen. First, a stoichiometric mixture of high-purity $Nd_2O_3$ (Aladdin, purity: 99.99%) and NiO (Aladdin, purity: 99.9%) were weighed and grounded thoroughly in an agate mortar under argon atmosphere. Then we pressed the mixture into pellets and sintered the pellets at 1000 °C for 24 hours. At last, the sintered pellets were reground, pelletized, and sintered at 1100 °C for one week with several intermediate regrinding and pressing processes, and eventually the single phase $Nd_4Ni_3O_{10}$ samples are obtained. Note that all sintering processes were performed in an oxygen atmosphere with an oxygen pressure of about 0.1 MPa. The samples of $Nd_4Ni_3O_8$ were prepared via a topochemical reduction process from $Nd_4Ni_3O_{10}$ using $CaH_2$ as a reducing agent. Three molar excess of $CaH_2$ and one molar of $Nd_4Ni_3O_{10}$ were grounded and pelletized in a glove box, and sealed in an evacuated quartz tube. The tube was slowly heated to 280 °C and maintained for 20 hours. Then the final phase of $Nd_4Ni_3O_8$ was obtained by washing the CaO byproduct and the residual $CaH_2$ out of the samples with saturated $NH_4Cl$ in anhydrous ethanol.

The room temperature powder x-ray diffraction patterns were collected using the Bruker *D8* Advance diffractometer with Cu-$K_\alpha$ radiation. The scan rate is 0.01°/step in the range of $2\theta$ from 10° to 90°. Rietveld refinement fitting [41] was performed by using the *TOPAS 4.2* software [42]. The magnetic properties were measured with a SQUID-VSM-7T (Quantum Design). The electronic resistivity was measured with a Physical Property Measurement System (PPMS, Quantum Design), and a standard four-probe method was used for the ambient pressure resistivity measurement. The Hall resistivity was measured using a six-probe method. The high-pressure electronic transports were measured via a four-probe van der Pauw method [43] in a diamond anvil cell (*cryo*-DACPPMS, Almax easyLab), and the ruby fluorescence method is used to determine the values of pressure [44].

# 3 Results and discussion

## 3.1 Sample characterization

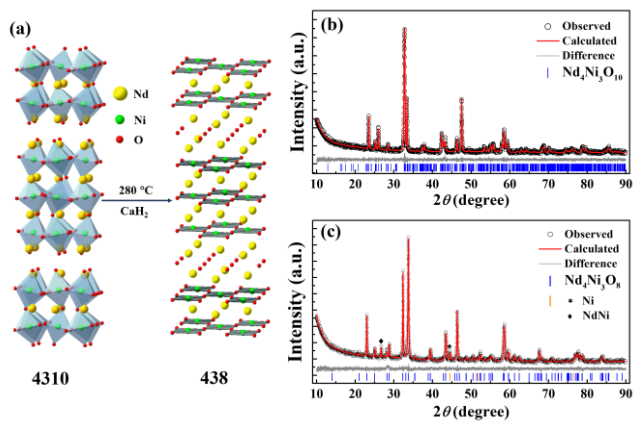

**Figure 1** (color online) (a) The crystal structures of Nd$_4$Ni$_3$O$_{10}$ and Nd$_4$Ni$_3$O$_8$ with the reduction procedure. (b, c) Powder x-ray diffraction patterns (black circles) and Rietveld fitting curves (Red lines) of Nd$_4$Ni$_3$O$_{10}$ and Nd$_4$Ni$_3$O$_8$ at room temperature.

**Table 1** Crystallographic data of Nd$_4$Ni$_3$O$_{10}$ and Nd$_4$Ni$_3$O$_8$ at room temperature.

| Compound | Nd$_4$Ni$_3$O$_{10}$ | Nd$_4$Ni$_3$O$_8$ |
|---|---|---|
| space group | $P2_1/a$ | $I4/mmm$ |
| $a$ (Å) | 5.3614(2) | 3.9142(9) |
| $b$ (Å) | 5.4497(5) | 3.9142(9) |
| $c$ (Å) | 27.399(8) | 25.296(7) |
| $V$ (Å$^3$) | 800.56(8) | 387.58(2) |
| $\rho$ (g/cm$^3$) | 4.680 | 4.546 |
| $R_{wp}$ (%) | 4.92 | 4.45 |
| $R_p$ (%) | 3.86 | 3.50 |
| GOF | 1.04 | 1.06 |

Figure 1(a) illustrates the crystal structures of Nd4310 (left) and Nd438 (right). Nd4310 crystallizes in the monoclinic symmetry (space group: $P2_1/a$) with disordered NiO$_6$ octahedra in perovskite (NdNiO$_3$)$_3$ blocks and rock-salt Nd-O layers. The compound Nd438 can be obtained by removing the oxygen atoms from the Nd-O layers in the vertex of NiO$_6$ octahedra and performing structural rearrangement from *RP* type Nd4310, and crystallizes in the tetragonal structure (space group: $I4/mmm$) [34]. The structure can be described as intergrowth of three corner-sharing square NiO$_2$ planes and R$_2$O$_2$ fluorite-type layers, stacking along the $c$ axis. The powder XRD data and relevant Rietveld refinements of Nd4310 and Nd438 are shown in Fig. 1(b, c). The purity of the Nd4310 and Nd438 phases are confirmed by indexing of all the observed diffraction peaks from XRD data. Two tiny impurity peaks which may arise from segregation phases nickel and NdNi alloy are observed in Nd438, which are shown in Fig. 1(c). The impurity phases in Nd438 should origin from the decomposition of Nd4310 which is a common feature in the low temperature reduction process [16, 45]. The results of Rietveld fitting and cell parameters of both samples are summarized in Table 1, which are in a good agreement with those in previous reports [28, 34, 46]. All of these show that our samples are of good quality.

**Table 2** BVS calculations of Nd and Ni in Nd$_4$Ni$_3$O$_{10}$ and Nd$_4$Ni$_3$O$_8$ at room temperature (Length means the bond length of Ni-O or Nd-O).

| Nd$_4$Ni$_3$O$_{10}$ | | Nd$_4$Ni$_3$O$_8$ | | |
|---|---|---|---|---|
| Elements | BVS | Elements | Length(Å) | BVS |
| Ni1 | 2.64 | Ni1 | 1.9571×4 | 1.76 |
| Ni2 | 2.79 | | | |
| Ni3 | 2.65 | Ni2 | 1.9583×4 | 1.76 |
| Ni4 | 2.35 | | | |
| Nd1 | 2.38 | Nd1 | 2.4862×4 | 2.56 |
| Nd2 | 2.85 | | 2.5686×4 | |
| Nd3 | 2.50 | Nd2 | 2.7247×4 | 3.03 |
| Nd4 | 3.28 | | 2.3125×4 | |

During the reduction process, the apical oxygen atoms of NiO$_6$ octahedra in Nd4310 have been removed or shifted. The migration of apical oxygen may cause the change of the valence state of Ni in Nd438. Thus, we perform bound valence calculations to get more information about the cationic valence or possible charge ordering state. The bound valence sum (BVS) values [47] can be calculated from the equation: $V = \sum exp[(r_0 - r_i)/B]$, where $r_i$ is the bond length, $r_0$ is 1.654 for Ni$^{2+}$- O bonds and 2.105 for Nd$^{3+}$- O bonds, and $B$ is an empirical parameter which is usually determined to be 0.37. The calculated BVS values of Nd4310 and Nd438 are summarized in Table 2. The BVS results yield the averaged oxidation state of Ni is +2.57 in Nd4310, which is close to the expected value (+2.6) and consistent with the previous report [28]. However, the averaged oxidation state Ni$^{1.76+}$ in Nd438 is higher than its expected value (+1.33, assuming Nd is +3). Noting that from the literature [36], two different types of defects have been observed in Nd438 through the transmission electron microscopy (TEM) study, which may lead to a slight increase of valence state of Ni in Nd438. Meanwhile, the BVS calculations on the rare earth cation of both materials (2.75 and 2.79 for Nd4310 and Nd438, respectively) reveal the average oxidation states of Nd are slightly lower than the formal oxidation state +3. The relatively low valence state for Nd seems to be a common feature especially in the nickelates with infinite-layer structure. The origin of such deviation may be attributed to the development of bonding character upon reduction or strain related effect as proposed in ref [34]. By the way, we do not find any indications for charge ordering in both Nd4310 and Nd438 from room temperature BVS calculations.

## 3.2 Magnetic and electrical transport properties

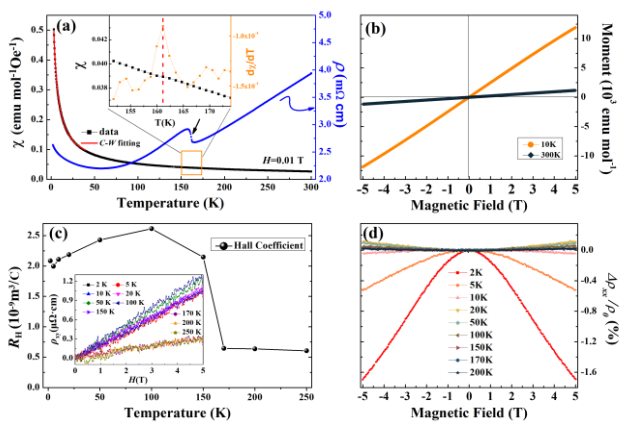

**Figure 2** (Color online) (a) Magnetic susceptibility and resistivity of $Nd_4Ni_3O_{10}$ as a function of temperature. Inset shows the enlarged view of $\chi$ - $T$ curve and the corresponding $d\chi/dT$ curve from 150 K to 175 K. (b) Magnetization hysteresis loops of $Nd_4Ni_3O_{10}$ at 10 K and 300 K. (c) Hall coefficient $R_H$ of $Nd_4Ni_3O_{10}$ as a function of temperature. Inset shows the $\rho_{xy}$ - $H$ curves at different temperatures. (d) Magnetoresistance $\Delta\rho/\rho_0$ as a function of magnetic field at different temperatures.

In Fig. 2(a), we show the data of magnetization and resistivity of Nd4310 to characterize the phase transition at about 162 K. As we can see, the temperature dependence of magnetic susceptibility and resistivity curves indicate a paramagnetic metallic behavior in a wide temperature range and a transition-related anomaly at about 162 K in $\rho$ - $T$ curve (as shown by the black arrow), which is consistent with previous reports [23, 28]. It is worth noting that a magnetic susceptibility kink consistent with the transition in $\rho$ - $T$ curve can also be detected from $d\chi/dT$ - $T$ curve as shown in the inset of Fig. 2(a). We perform the magnetization hysteresis ($M$ - $H$) measurements for Nd4310 at temperatures of 10 K and 300 K, and present the data in Fig. 2(b). The $M$ - $H$ curves show linear dependence and the magnetization is unsaturated up to $H$ = 5 T. The results indicate that the title compound is paramagnetism in nature and do not contain the ferromagnetic component in the measured temperature range. Then we fit the temperature dependence of the magnetic susceptibility curve by the Curie-Weiss ($C$ - $W$) law $\chi = \chi_0 + C/(T + T_\theta)$ in the low temperature (3 - 40 K) region. The fitting values of Pauli paramagnetism $\chi_0$ and Curie constant $C$ are 0.0349 emu·mol$^{-1}$Oe$^{-1}$ and 2.62 emu·K·mol$^{-1}$Oe$^{-1}$, respectively. And the effective magnetic moment ($\mu_{eff}$) determined through the fitting is about 4.58 $\mu_B$/f.u. in Nd4310.

To get a deeper understanding of the density-wave-like transition around 162 K, we show the data of Hall coefficient $R_H$ of Nd4310 at various temperatures in Fig. 2(c). The inset of Fig. 2(c) shows the raw data of the transverse resistivity $\rho_{xy}$ at different temperatures. The linear positive values of $\rho_{xy}$ indicate hole type carriers in whole temperature range. Furthermore, the values of Hall coefficient $R_H$ around the temperature about 162 K changes significantly. For temperatures below 162 K, the clear temperature dependence of Hall coefficient may be attributed to the multi-band conduction of the charge carriers. The almost constant Hall coefficient (about 6 × 10$^{-10}$m$^3$C$^{-1}$) above 162 K may correspond to a dominating single band property. Taking all the experimental evidence about the phase transition $T^*$ into consideration, including the step-like changes of the cell parameters and specific heat peaks reported by Li $et\ al.$ [28], the transition around 162 K should be a first-order transition with possible reconstruction of the Fermi surface. The transition in Nd4310 might origin from intertwined spin and charge-density waves, similar to the case in $La_4Ni_3O_{10}$ [48]. Fig. 2(d) shows the field dependence of $MR$ for Nd4310 at different temperatures. With the application of magnetic field, a negative $MR$ is visible below 10 K. The negative $MR$ increases with decreasing temperature and reaches its maximum $\Delta\rho/\rho_0$ = -1.7% at 2 K with $H$ = 5 T. The sensitive magnetic field dependence of negative $MR$ may be due to the delocalization or suppression of the Kondo-like scattering, which can lead to the decrease of resistivity under magnetic field. These pictures are basically consistent with the resistivity upturn below 50 K in our $\rho$ – $T$ data. Noting that the low-temperature negative $MR$ behaviors in present material are different from the reported $La_4Ni_3O_{10}$ and $Pr_4Ni_3O_{10}$ [26, 27], in which a positive $MR$ is observed below 10 K. Thus, it is possible that the $MR$ behavior in these materials are associated with the magnetism of $Nd^{3+}$ or oxygen nonstoichiometry.

For comparison, we present the temperature dependence of magnetization curves for Nd438 under different magnetic fields (1 T and 3 T) in Fig. 3(a). The magnetization increases with decreasing temperature and demonstrates a paramagnetic behavior in the whole temperature region. Combining the magnetization hysteresis curves measured at different temperatures as shown in Fig. 3(b), we can separate a ferromagnetic component from the paramagnetic background which should come from the ferromagnetic impurity of nickel. Then we use a similar method as described in the previous report [16] to derive the inherent magnetization of Nd438 by subtracting the $M$ - $T$ curve of 1 T from that of 3 T. The derived curve is presented in the inset of Fig. 3(a). The $C$ - $W$ like temperature dependence of susceptibility curve is similar to its parent compound Nd4310, but different from the brother compound $La_4Ni_3O_8$ in which a field dependent sharp drop of magnetization at low temperatures is observed. We also use the $C$ - $W$ law to fit the $\chi_P - T$ curve in low temperature region (3 - 40 K) to analyze the magnetic behavior of Nd438. The fitting values of Pauli paramagnetism $\chi_0$ and Curie parameter $C$ are 0.0466 emu·mol$^{-1}$Oe$^{-1}$ and 1.95 emu·K·mol$^{-1}$Oe$^{-1}$, respectively. And the effective magnetic moment ($\mu_{eff}$) determined through the fitting is about 3.95 $\mu_B$/f.u. in Nd438. The fitting result of Nd438 is similar to that of parent compound Nd4310. The relatively large value of $\mu_{eff}$ may suggest that the magnetic moments in both systems are contributed by the Nd ions.

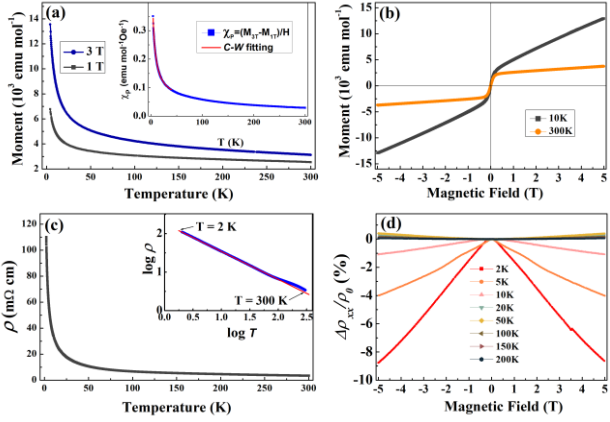

**Figure 3** (Color online) (a) Magnetic moment of $Nd_4Ni_3O_8$ as a function of temperature with the applied fields of 1 T and 3 T. Inset shows the curve of $\chi_P$ - $T$ curve, where $\chi_P$ is obtained by subtracting the curve of 1 T from that of 3 T. (b) Magnetization hysteresis loops of $Nd_4Ni_3O_8$ at 10 K and 300 K. (c) Resistivity of $Nd_4Ni_3O_8$ as a function of temperature. Inset shows the curve of logarithmic resistivity as a function of logarithmic temperature. (d) Magnetoresistance $\Delta\rho/\rho_0$ as a function of magnetic field at different temperatures.

Figure 3(c) presents the resistivity of Nd438 as a function of temperature at zero field. The result reveals an insulating behavior in the temperature range from 2 K to 300 K, with a room temperature resistivity about 3.4 $\Omega$ cm. Noting that the absolute resistivity of Nd438 is similar to $Pr_4Ni_3O_8$ but the electrical transport behaviors are very different. For $Pr_4Ni_3O_8$, the $\rho$ - $T$ curve exhibits metallic behavior in the temperature range of 2 - 300 K [32]. Then we try to fit the electrical transport data to analyze the conduction type of the insulating behavior. Neither band gap model nor variable range hopping model could be used to fit the insulating behavior. Surprisingly, the transport behavior can be well described by a $\log\rho \propto -\log T$ relationship in the whole temperature region we measured. To our knowledge, the linear $\log\rho \propto -\log T$ behavior can be classified as the multi-phonon hopping (MPH) conduction model [49, 50], which indicates that carriers are coupled with both acoustic and optical phonons. As for the origin of intrinsic insulating behavior in Nd438, it remains unclear yet. The band structure calculations indicate that the material should be a metal [12], but the insulating behavior may be attributed to the possible buckling of the $NiO_2$ square-planar or a hidden antiferromagnetic order inducing band-gap opening at the Fermi level, similar to the case in $La_4Ni_3O_8$ [39, 51]. Moreover, the robust insulating behavior may also be caused by the topotactic intercalation of hydrogen in Nd438, in which the phase of $Nd_4Ni_3O_8H_x$ may be formed during the $CaH_2$ reduction reaction. The field dependence of $MR$ for Nd438 is shown in Fig. 3(d). It is evident that the value of $MR$ for Nd438 at 2 K (about -9% at 5 T) is much larger than that of Nd4310. The larger negative $MR$ in present material may be caused by the de-localization of the low-dimensional correlated oxides or Kondo-like scattering.

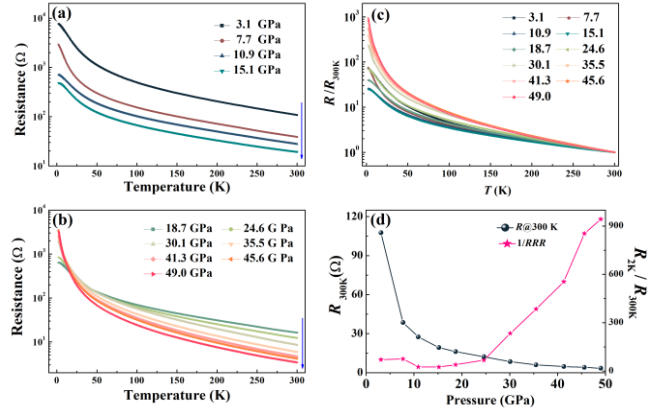

**Figure 4** (Color online) (a, b) Electrical resistance of $Nd_4Ni_3O_8$ as a function of temperature at various pressures up to 49 GPa. (c) The normalized resistance as a function of temperature of $Nd_4Ni_3O_8$ at various pressures. (d) The plots of resistance at room temperature ($R_{300K}$) and $R_{2K}/R_{300K}$ of $Nd_4Ni_3O_8$ as a function of applied pressure.

Figure 4(a-c) display the resistance and normalized resistance ($R/R_{300K}$) in semi-logarithmic scale as a function of temperature of Nd438 sample from 2 K to 300 K at various pressures. The absolute resistance values at room temperature ($R_{300K}$) decrease continuously with the increase of applied pressure up to 49 GPa as shown by the blue arrowed line in Fig. 4(a, b). And the $R_{300K}$ as a function of pressure are plotted in Fig. 4(d). For pressures below 15.1 GPa, the low temperature resistance ($R_{2K}$) also decreases with increasing pressure. The pressure-induced enhancement in conductivity may be due to the modification of the grain boundaries or the narrowing of hopping distance under compression. When the pressure is increased further (above 18.7 GPa), the insulating behavior is still existing and even stronger. Carefully checking the resistance at low temperatures, a crossover of resistance values at different pressures is observed below 50 K. The abnormal upturn trend of resistance at low temperature may indicate that the insulating behavior becomes much stronger when applying a high pressure. We present the curves of normalized resistance versus temperature at various pressures in Fig. 4(c) and summarize $R_{2K}/R_{300K}$ versus pressure curve in Fig. 4(d). One can see that, the values of $R_{2K}/R_{300K}$ first decrease slightly with the increase of pressure and then increase sharply. At the highest pressure (49.0 GPa) reached in our measurement, the compound Nd438 becomes a strong insulator with the ratio $R_{2K}/R_{300K}$ of about 943. The enhanced insulating behavior of the sample under high pressure may be attributed to the pressure-induced lattice distortions, which is particularly the case for materials with relatively strong spin-orbit interactions. Therefore, our high pressure resistance measurements reveal the enhancement of insulating behavior in Nd438. However, superconductivity has not been observed in Nd438 under pressures up to 49 GPa.

## 4 Conclusions

To summarize, we successfully synthesize the trilayer nickelates $Nd_4Ni_3O_{10}$ and $Nd_4Ni_3O_8$, and conduct detailed experimental investigations through magnetization and resistivity measurements. The magnetization and resistivity data for $Nd_4Ni_3O_{10}$ indicate paramagnetic metal behavior with a phase transition at about 162 K. A hole-type feature of charge carriers with obvious change of $R_H$ around $T^*$ is observed from Hall resistivity measurement. The significant reduced carrier concentration and strong temperature dependence of $R_H$ below $T^*$ indicate a possible reconstruction of the Fermi surface around $T^*$. By using a low temperature topochemical reduction method, we obtained the compound $Nd_4Ni_3O_8$ with square $NiO_2$ planes. The magnetization and electrical transport measurements reveal a Curie-Weiss like paramagnetic feature and an insulating behavior. The high-pressure study on electrical resistance of Nd438 shows the robust insulating behavior, and no superconductivity is observed up to 49 GPa above 2 K.

*This work was supported by the National Key R&D Program of China (Grant No. 2016YFA0300401, 2016YFA0401704), National Natural Science Foundation of China (Grant No. A0402/11534005 and A0402/11674164), and the Strategic Priority Research Program of Chinese Academy of Sciences (Grant No. XDB25000000).*